\documentstyle[12pt,epsf]{article}
\begin {document}
~\hspace*{7.1cm} ADP-AT-96-9 (Revised)\\
~\hspace*{8cm} J. Phys. G: Nucl. Part. Phys., submitted\\[2cm]

\begin{center}
{\large \bf Cosmic-Ray Electrons and the Diffuse Gamma-Ray Spectrum}
\\[1cm]
T.A. Porter and R.J. Protheroe \\
	Department of Physics and Mathematical Physics \\
	University of Adelaide, Adelaide 5005, Australia\\[2cm]
\end{center}

\noindent Short Title: Cosmic-ray electrons and diffuse gamma-ray 
spectrum\\[2cm]

         interactions)

95.85.Pw Gamma-ray

98.70.Vc Background radiations

\newpage

\begin{center}
\underline{Abstract}\\
\end{center}
The bulk of the diffuse galactic gamma-ray emission above a few
tens of GeV has been conventionally ascribed to the decay of
neutral pions produced in cosmic-ray interactions with
interstellar matter.  Cosmic-ray electrons may, however, make a
significant contribution to the gamma-ray spectrum at high
energies, and even dominate at TeV--PeV energies depending on
their injection spectral index and acceleration cut-off energy.
If the injection spectrum is flat, the highest energy electrons
will also contribute a diffuse hard X-ray/soft gamma-ray flux via
synchrotron emission, and this may offer an explanation for the
OSSE observation of a steep spectrum below a few MeV from the
inner Galaxy.  We perform a propagation calculation for 
cosmic-ray electrons, and use the resulting interstellar electron
spectrum to obtain the gamma-ray spectrum due to inverse Compton,
synchrotron and bremsstrahlung interactions consistently from MeV
to PeV energies.  We compare our results with available
observations from satellite-borne telescopes, optical $\check{\rm
C}$erenkov telescopes and air shower arrays and place constraints
on the injection spectrum of cosmic-ray electrons.  With future
observations at TeV--PeV energies it should be possible to
determine the average interstellar spectrum of cosmic-ray
electrons, and hence estimate their spectrum on acceleration.

\section{Introduction}
\label{secIntro}

Cosmic-rays with energies up to $\sim 100$ TeV are thought to be accelerated
by the 1st order Fermi mechanism at supernova shocks 
(see Jones and Ellison \cite{Jones91A} for a recent review), and recently
the EGRET instrument on the Compton Gamma-Ray Observatory has detected 
gamma-ray signals above 100 MeV from at least two 
supernova remnants (SNR) $-$ IC 443 and $\gamma$ Cygni \cite{Esposito96A}.
Further evidence for particle acceleration comes from recent ASCA 
observations of non-thermal X-ray emission from SN 1006 \cite{Koyama95A}, 
and correlation of ASCA and ROSAT observations of non-thermal 
X-ray emission from IC 443 \cite{Keohane97A}. 
Reynolds \cite{Reynolds96A}
and Mastichiadis \cite{Mastichiadis96A} interpret the former as synchrotron
emission by electrons accelerated in the remnant up to energies as high
as 100 TeV, while Keohane {\em et al} \cite{Keohane97A} argue that the 
latter is due to synchrotron emission by electrons accelerated to $\sim 10$
TeV.
Recently Pohl \cite{Pohl96A} has suggested electron acceleration up to 100 TeV 
energies may not be unique to SN 1006, and that other 
acceleration sites of high energy electrons probably exist in the Galaxy.

Electrons accelerated to 100 TeV energies would eventually escape their
acceleration sites and diffuse in the Galaxy, cooling through synchrotron 
radiation and inverse Compton (IC) scattering on the galactic magnetic 
and radiation fields respectively.
For synchrotron cooling, electrons of these energies in a magnetic field 
strength $\sim 6$ $\mu$G would 
give a diffuse flux of radiation in the X-ray regime, while IC scattering
of 100 TeV energy electrons on the cosmic microwave background radiation
(CMBR) would give a diffuse flux of gamma-rays at TeV energies.
The spectrum of radiation produced by these processes is dependent on the
high energy interstellar electron spectrum, which in turn is 
dependent on the initial source spectrum, distribution of sources
and propagation.
If this radiation is detectable it would provide a means of estimating 
the average interstellar electron spectrum, and hence the spectrum of 
electrons at acceleration.

Protheroe and Wolfendale \cite{Protheroe80A}, in an approximate 
calculation, have considered the dual role of ultrarelativistic electrons
in producing the diffuse galactic radiation from hard X-rays to TeV 
energies and above, and an analysis of Uhuru data by Protheroe 
{\it et al} \cite{Protheroe80B} has indicated a considerable contribution 
by synchrotron emission in the soft X-ray band. 
However, more recent calculations of the diffuse gamma-ray spectrum
\cite{Bertsch93A,Strong95B,Strong96B,Skibo96A,Hunter97A} have generally
neglected synchrotron emission as a significant production process.
Inverse Compton scattering on the 
ambient galactic photon fields, on the other hand, is recognised as a 
important contributor to the diffuse gamma-ray spectrum at MeV to GeV
energies.
In particular, detailed predictions of IC gamma-rays above $70 - 100$ MeV 
have been made by Bloemen \cite{Bloemen85A} and Chi {\it et al} \cite{Chi89A},
with estimates of the contribution by this component being up to 
$\sim 50 \%$ of the diffuse intensity at medium galactic latitudes 
($|b| = 10^\circ$ to $20^\circ$). 
An analysis of EGRET data by Giller {\it et al} \cite{Giller95A} has 
suggested a contribution by IC of $\sim 30 \%$
for medium latitudes, and up to $\sim 45 \%$ toward the galactic pole, in 
the energy range $30 - 4000$ MeV.
Strong {\it et al} \cite{Strong96B} have shown that an IC component
is required to provide a good fit to the gamma-ray data from 1 MeV to 1 GeV;
see also Bertsch {\em et al} \cite{Bertsch93A} and Hunter {\em et al}
\cite{Hunter97A}.

In this paper we consider possible injection spectra of primary
cosmic-ray electrons, and the resulting diffuse gamma-ray
spectrum of the Galaxy.
Starting with a power-law spectrum of electrons at acceleration, we 
propagate electrons in the Galaxy using a diffusion model which is consistent
with the observed cosmic-ray secondary to primary data and $^{10}$Be 
abundance to obtain the interstellar electron spectrum.
Realistic models of the galactic matter distribution, magnetic field and
interstellar radiation field (ISRF) are used in the propagation calculation.
Inverse Compton, bremsstrahlung and synchrotron production spectra are 
calculated using the electron spectra resulting from the propagation 
calculation, and we give a consistent treatment of high energy photon
production by these processes from keV to PeV energies.
Our predictions are then compared with satellite observations at keV to 
GeV energies 
\cite{Strong96B,Strong88A,Strong94A,Purcell95A,Purcell96A,Strong96A}, optical 
$\check{\rm C}$erenkov telescope observations at TeV energies 
\cite{Reynolds93B} and air shower observations 
\cite{Matthews91A,Karle95A,Borione97A} at $50 - 1000$ TeV energies.

In Section \ref{sectionPropCalc} we describe an efficient Monte Carlo method 
that allows us to vary source distributions, or injection spectra, without 
having to repeat the propagation calculation for each different case.
We then use the method to obtain the interstellar electron spectrum
for representative regions of the Galaxy.
We consider various injection spectral indices, and constrain the model
spectra using direct measurements of the local electron spectrum and the 
galactic non-thermal radio emission.
Electron spectra that satisfy the constraints are then used in Section 
\ref{sectionGammaSpec} to calculate production spectra for IC,
bremsstrahlung and synchrotron emission processes, and to obtain the 
expected gamma-ray intensity.
In Section \ref{sectionDiscuss} we discuss the limitations of our model, and
the implications of our predictions.

\section{The Electron Propagation Calculation}
\label{sectionPropCalc}

\subsection{Numerical Method}
\label{subsectionNumMeth}
The standard transport equation for cosmic-ray electrons undergoing continuous
energy losses is \cite{Ginzburg64A}

\begin{equation}
\frac{\partial n\left( E, \vec{r}, t \right)}{\partial t} =
q\left(E, \vec{r} \right) + K(E) \nabla^2 n\left( E,\vec{r} ,t\right)  + 
\frac{\partial}{\partial E} 
\left[ n\left( E,\vec{r},t \right)\frac{d E}{d t}
\right]
\label{equCRTrans}
\end{equation}

\noindent
where $n\left(E , \vec{r} , t \right)$ is the number density of electrons,
$q\left( E, \vec{r} \right)$ is the source function of 
cosmic-ray electrons assuming a constant injection rate, the second term on 
the right represents energy 
dependent spatial diffusion of the electrons with scalar diffusion 
coefficient $K(E)$ and the third term represents 
the continuous energy losses of the electrons. We seek steady state
solutions $n\left( E, \vec{r} \right)$ and solve Equation \ref{equCRTrans} by 
Monte Carlo methods which we describe below. 

Consider electrons of energy $E'$ released at $\vec{r}\,'$ to diffuse
through the Galaxy and continuously lose energy by interacting with
the background radiation, magnetic field and matter.
We define $p(E,\vec{r}; E',\vec{r}\,')$ to be the probability density 
(cm$^{-3}$)
at $\vec{r}$ of points in space at which the energy of the electrons
was precisely $E$.
Given the energy-loss rate

\begin{equation}
{dE \over dt} = -b(E,\vec{r}) \hspace{5mm} {\rm (GeV \; s^{-1})},
\label{equNumMeth1}
\end{equation}

\noindent
the average time spent with energy between $E$ and $(E + dE)$ is
$dE/b(E,\vec{r})$ if the electron is located at or close to $\vec{r}$. 
Hence, the average time spent 
by an electron with energies between $E$ and $(E + dE)$ per unit
volume at $\vec{r}$ is $p(E,\vec{r}; E',\vec{r}\,') dE/b(E,\vec{r})$.
Thus, for some source distribution $q(E, \vec{r})$ 
(GeV$^{-1}$ cm$^{-3}$ s$^{-1}$) we obtain the number density of
electrons $n(E, \vec{r})$ (GeV$^{-1}$ cm$^{-3}$) at $\vec{r}$

\begin{equation}
n(E, \vec{r}) = b(E,\vec{r})^{-1} \int dV' 
\int_E^\infty dE' q(E', \vec{r}\,') p(E,\vec{r}; E',\vec{r}\,').
\label{equNumMeth2}
\end{equation}

\noindent
The most time-consuming part of the calculation is working out
$p(E,\vec{r}; E',\vec{r}\,')$ because this contains all the information
about propagation, energy losses and interactions during propagation.

We make the approximation that propagation takes place only in the 
$z$ direction (perpendicular to the galactic plane), and use a
two-dimensional model of the radiation field, magnetic field,
and matter density, in which these quantities depend on galactocentric 
radius, $R$, and height above the plane, $z$. For this case

\begin{equation}
n(E, R,z) \approx b(E,R,z)^{-1} \int dz' 
\int_E^\infty dE' q(E, R,z') p(E,R,z; E',R,z')
\label{equNumMeth3}
\end{equation}

\noindent
where the $R$ dependence has been retained due to the two-dimensional 
dependence of the matter distribution, radiation and magnetic fields.
The probability density $p(E,R,z; E',R,z')$, which we shall
refer to as the ``probability matrix'', is essentially the Green's function 
for Equation \ref{equCRTrans} and is calculated by the 
Monte Carlo method as described later. 
We can re-use $p(E,R,z; E',R,z')$ to obtain $n(E, R,z)$ for different
source spectra or distributions $q(E, R,z)$ simply by
performing the integrals over energy and volume given above.
This is particularly useful if we wish to consider different 
cosmic-ray electron source spectra.

To demonstrate the reliability of our method we calculate $n(E,R,z)$ 
for two simple cases, and compare with the analytical solutions. 
We consider: (i) the solution of Equation \ref{equCRTrans} with energy 
independent diffusion coefficient $K$ and constant energy loss rate
$b(E,R,z) = \alpha$, and (ii) the solution of Equation \ref{equCRTrans}
with energy dependent diffusion coefficient $K(E) = K (E/E_0)^\delta$
and energy loss rate $b(E,R,z) = B E^2$.
The motivation for this is that cases (i) and (ii) approximate the 
diffusion and energy loss mechanisms of cosmic-ray electrons at low and 
high energies respectively.
For simplicity we obtain the solution $n(E,R,z)$ for both cases without
boundaries.

The Green's function for case (i) is \cite{Porter97A}

\begin{equation}
G_{(i)} (E,R,z;E',R,z') = \frac{q_0}
{\sqrt{4 \pi K \alpha \left| E - E' \right|}}
\exp \left( - \frac{\alpha \left( z - z' \right)^2}{4 K \left| E - E' \right|}
\right)
\label{equGFNCase1}
\end{equation}

\noindent
and for case (ii)

\begin{equation}
G_{(ii)} (E,R,z;E',R,z') = \frac{q_0}{\sqrt{4 \pi B^2 E^4 
(\lambda (E) - \lambda (E')
)}} \exp \left( -\frac{(z - z')^2}{4(\lambda (E) - \lambda (E'))} \right)
\label{equGFNCase2}
\end{equation}

\noindent
where $\lambda (E) = - \int_E ^\infty K(X)/b(X,R,z) dX$. By definition,  
the Green's functions are solutions of Equation \ref{equCRTrans} for a 
source function $q(E,z) = q_0 \delta(E - E') \delta(z - z')$ 
(GeV$^{-1}$ cm$^{-3}$ s$^{-1}$). 

\begin{figure}[htb]
\vspace{17.0cm}
\includegraphics{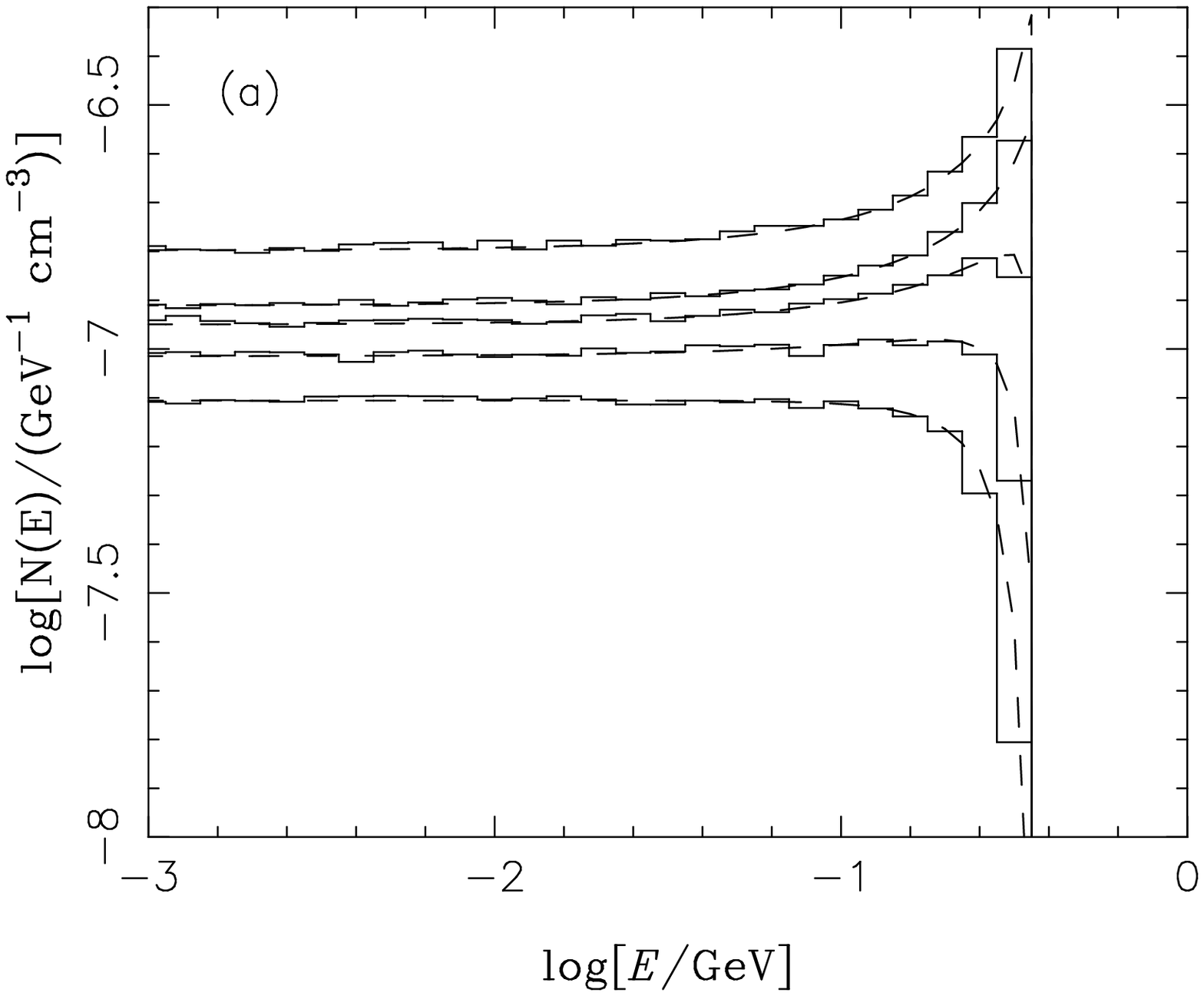}
\includegraphics{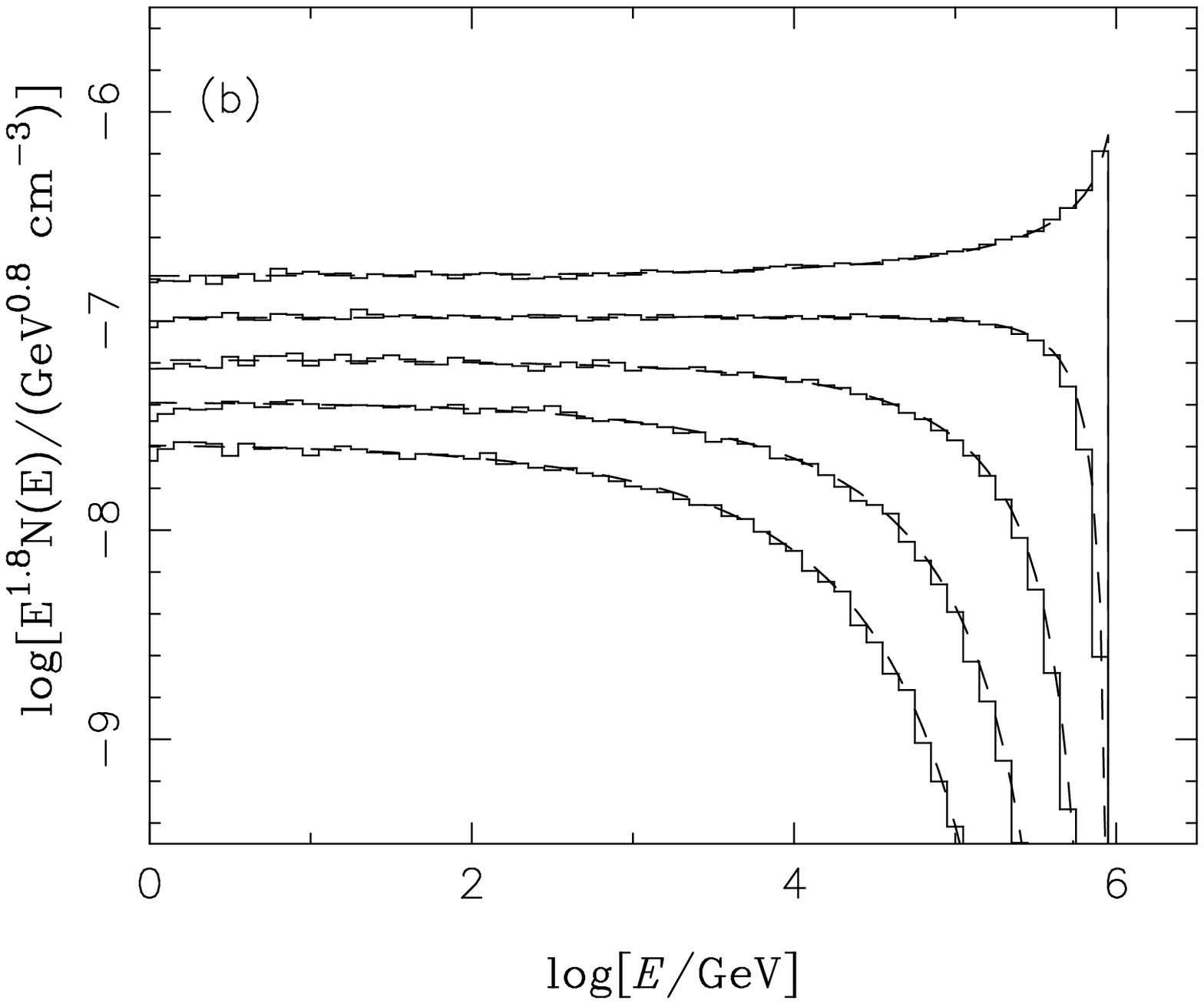}
\caption{Particle distribution in $E$ and $z$ calculated using (a) Equation 
\protect\ref{equGFNCase1} and the numerical method  
for particles released at $z = 0$ and initial energy $E' = 10^{-0.4}$ GeV, 
and (b) Equation \protect\ref{equGFNCase2} and the 
numerical method for particles released at $z = 0$ kpc and initial energy
$E' = 10^6$ GeV. 
In each case the analytical solutions are plotted as dashed
lines and the Monte Carlo results as histograms.
The upper curve 
shows the distribution at $z = 0$ kpc with lower curves showing the 
distribution for $z = $ 0.5, 1.0, 1.5, 2.0 kpc.
To clarify the diagrams the upper curve in case (a) has been offset a 
factor $10^{0.1}$ relative to the other curves while in case (b) the 
lower curves have been offset relative to the $z = 0$ kpc curve by a 
factor $10^{-0.2}$, 10$^{-0.4}$ etc.}
\label{figGFNComp}
\end{figure}

In the Monte Carlo method, for each case we obtain $p(E,R,z;E',R,z')$
with the same source function.
For case (i) we take 41 energy bins at intervals of $\Delta \log E = 0.1$ 
with mid-bin energy starting at $10^{-3}$ GeV.
The Monte Carlo procedure is as follows.
A particle is injected at $z = 0$ with initial energy $E_j$.
Based on random walk theory and its relation to diffusion 
\cite{Chandrasekhar43A}, the diffusion in $z$ is simulated by multiplying
a randomly sampled normal deviate, $\zeta$, by the standard deviation

\begin{equation}
\sigma_z = {\rm min} \left( \sqrt{2 K \Delta t}, \sigma_{z_{max}} \right)
\label{equMCStepSize}
\end{equation}

\noindent
where $\Delta t = \Delta E/b(\sqrt{E_j E_{j-1}},R,z)$ with 
$\Delta E = (E_j - E_{j-1})$ being the difference between the current 
mid-bin energy
and the next lowest mid-bin energy, and $\sigma_{z_{max}}$ is the maximum 
standard deviation and is chosen to minimise computing time while 
ensuring the Monte Carlo and analytical results agree; 
$\sigma_{z_{max}}$ is chosen to be 
small compared to the distance over which physical parameters of the model
change significantly.
The new position is obtained by adding $\Delta z = \zeta \sigma_z$ to the 
current position. 
If $\sigma_z < \sigma_{z_{max}}$ the particle's energy is set to $E_{j-1}$
and, if the new position is in the ``observing region'' i.e. the region
for which we wish to obtain $n(E,R,z)$, the particle is 
recorded as having been observed with energy $E_{j-1}$.
Otherwise, if $\sigma_z = \sigma_{z_{max}}$, the energy lost by the particle 
while diffusing is calculated and subtracted from its current energy. 
If the particle's energy falls below $E_{j-1}$, the approximate 
position where its 
energy became lower than $E_{j-1}$ is determined and, if this position is
within the observing region, the particle is recorded as having been observed
with energy $E_{j-1}$. 
The energy bin counter $j-1$ is then decremented by 1, and 
the above procedure is repeated until the particle reaches some 
large distance away from the galactic plane, taken to be 20 kpc in this 
instance; particles diffusing out to this distance would have lost an 
amount of energy large enough to place them below the lowest particle
energy in the simulation, hence they are no longer of interest. 
The probability matrix is computed by injecting $N$ particles
in each of the source energy bins and following the Monte Carlo 
procedure outlined above, the final result being divided by $N$ and the 
volume of the observing region. 
Figure \ref{figGFNComp}a compares the distribution of particles in 
$E$ and $z$ for source energies $E' = 10^{-0.4}$ GeV calculated using the 
above method for $N = 10^5$ particles and $\sigma_{z_{max}} = 50$ pc, and 
with Equation \ref{equGFNCase1} for $q_0 = 1$ cm$^{-2}$ s$^{-1}$, 
$K = 2.5 \times 10^{28}$ cm$^2$ s$^{-1}$ and $\alpha = 5 \times 10^{-16}$ GeV 
s$^{-1}$. 

For case (ii) we take 61 energy bins at intervals of $\Delta \log E = 0.1$ 
with mid-bin energy starting at $10^0$ GeV, and follow the Monte Carlo 
procedure outlined above.
Figure \ref{figGFNComp}b
compares the distribution of particles in $E$ and $z$ for source energies
$E' = 10^6$ GeV obtained from the probability matrix calculated using 
the Monte Carlo method with the analytical result (Equation \ref{equGFNCase2}) 
for $N = 10^5$ particles, $\sigma_{z_{max}} = 50$ pc, $q_0 = 1$ 
cm$^{-2}$ s$^{-1}$,  
$K = 2.5 \times 10^{28}$ cm$^2$ s$^{-1}$, $E_0 = 3$ GeV, $\delta = 0.6$ and
$B = 9.12 \times 10^{-17}$ GeV$^{-1}$ s$^{-1}$. 

In both cases the agreement between the Monte Carlo and analytical results 
is excellent. The particular choice of $\sigma_{z_{max}}$ used above was found
to give the best compromise, minimising computing 
time while maintaining good agreement with the analytical results. 
Making $\sigma_{z_{max}}$ much larger 
resulted in the particle distribution at low {\it observing} energies 
falling below the analytical results, while making it smaller ensured 
accuracy but produced unacceptably long run-times for the number of particles
chosen. Similarly simulations were performed for
varying $N$ and, in conjunction with our choice of $\sigma_{z_{max}}$, it was 
found $N = 10^5$ particles yielded the best compromise for computing time
versus statistical accuracy.

\subsection{Calculated Electron Spectra}
\label{subsectionCalc}
For the propagation calculation we adopt a halo half-height of 3 kpc 
\cite{Lukasiak94A}, and use a diffusion coefficient which is constant at
$2.5 \times 10^{28}$ cm$^2$ s$^{-1}$ below a magnetic rigidity of 
$\rho = 3$ GV and increases as $(\rho /3{\rm \; GV})^{0.6}$ above 3 GV 
\cite{Strong95A}. 
The modified source distribution of Webber {\it et al}
\cite{Webber92A} is used for the radial distribution of primary cosmic-ray 
electron sources. The sources are assumed to be uniformly distributed in $z$
with a half-height of 0.15 kpc comparable to the scale height of SNR 
\cite{Gaensler95A}. 

Theories of cosmic-ray acceleration predict a power-law dependence for the
injection spectrum of particles $q(E) \propto E^{-\gamma}$ \cite{Jones91A}. 
Studies of the non-thermal radio emission of the Galaxy give the following 
values for 
the power-law spectral index ($I_\nu \propto \nu^{-\alpha}$): 
$\alpha = 0.55 - 0.6$ at 5 to 80 MHz
in the direction of the galactic pole \cite{Webber80A}; 
$\alpha = 0.5 - 0.6$ at 38 to 408 MHz over the northern galactic
hemisphere \cite{Lawson87A};
$\alpha = 0.7$ for 408 to 5000 MHz for low latitudes toward the inner Galaxy 
\cite{Broadbent89A}; and $\alpha = 0.85$ in the plane, increasing to 
$\alpha = 1.0$ at higher latitudes for 408 to 1020 MHz \cite{Reich88A}.
Given the relationship $\alpha = (\gamma - 1)/2$ we see $\gamma = 2 - 3$ 
using the published ranges.
Due to the continuous energy losses suffered by electrons the source 
spectrum is not uniquely determined from the radio emission.
We therefore consider values of $\gamma$ at injection in the range 2 to 2.4.

Throughout this paper, we take the galactocentric 
radius of the Sun to be $R_S = 8.5$ kpc. 
Where the empirical models we use in our calculations use a different $R_S$
we scale them appropriately to our adopted value.
The maximum extent of the Galaxy in $R$ is taken to be 16 kpc.

For the ambient radiation fields of the Galaxy we use the model of Chi and
Wolfendale \cite{Chi91A} for the ultraviolet to near infra-red (`optical'),
the cold dust emission curve of Cox {\it et al} \cite{Cox86A}
for the far infra-red, and we take the temperature of the CMBR as 2.735 K. 
For electron interactions with matter we model three
components of the interstellar medium: the distribution of 
HI as given by Dickey and Lockman \cite{Dickey90A} with column density 
$\int n_{\rm HI} dz$ = $6.2 \times 10^{20}$ cm$^{-2}$ and a FWHM of 0.23 kpc
at $R_S$; the H$_2$ distribution of Bronfman {\it et al} with column 
density $\int n_{\rm H_2} dz$ = $3.8 \times 10^{20}$ cm$^{-2}$ and a 
FWHM of 0.14 kpc at $R_S$; and the HII distribution of Reynolds 
\cite{Reynolds90A,Reynolds91A,Reynolds93A} with column density 
$\int n_{\rm HII} dz$ = $2 \times 10^{20}$ cm$^{-2}$ and a FWHM of 3 kpc
at $R_S$. 
A contribution by helium consistent with the observed abundance is also
included.
The concentric ring model of Rand and Kulkarni \cite{Rand89A} is used for the
regular component of the galactic magnetic field and we adopt the value of
5 $\mu$G derived by these authors for the magnitude of the random component.
Energy loss formulae in both the non-relativistic and relativistic 
case for the various interactions electrons can undergo
with the interstellar medium, radiation and magnetic fields are taken 
from Ginzburg and Syrovatskii \cite{Ginzburg64A}; we consider ionisation 
loss and bremsstrahlung on both the neutral and ionised medium, and 
synchrotron and IC losses. 
For high energies the IC energy losses on the ISRF are in the
Klein-Nishina regime and we calculate the energy loss rate using Monte Carlo
methods as described by Protheroe \cite{Protheroe86A}.

We divide the Galaxy up into radial bins of half-width 1 kpc centred on 
$R$ = 1, 3, 5, 7, 9, 11, 13 and 15 kpc. 
The propagation parameters for 
each radial bin are then computed by averaging the models of the matter
distribution, radiation and magnetic fields over the inner radius to the 
outer radius of each bin. 
The electron spectra calculated for each 
bin are then taken to be representative of that region of the Galaxy.

To obtain the probability matrix for the $i$-th radial bin, we take 
111 energy bins at intervals of
$\Delta \log E = 0.1$ with mid-bin energy starting at $10^{-3}$ GeV. 
We inject particles uniformly within the source region and follow the 
Monte Carlo procedure outlined in the previous section. 
A value of $\sigma_{z_{max}}$ of 50 pc is used, and $N = 10^5$ particles are 
injected at each of the source energy bins.

\begin{figure}[htb]
\vspace{15.0cm}
\includegraphics{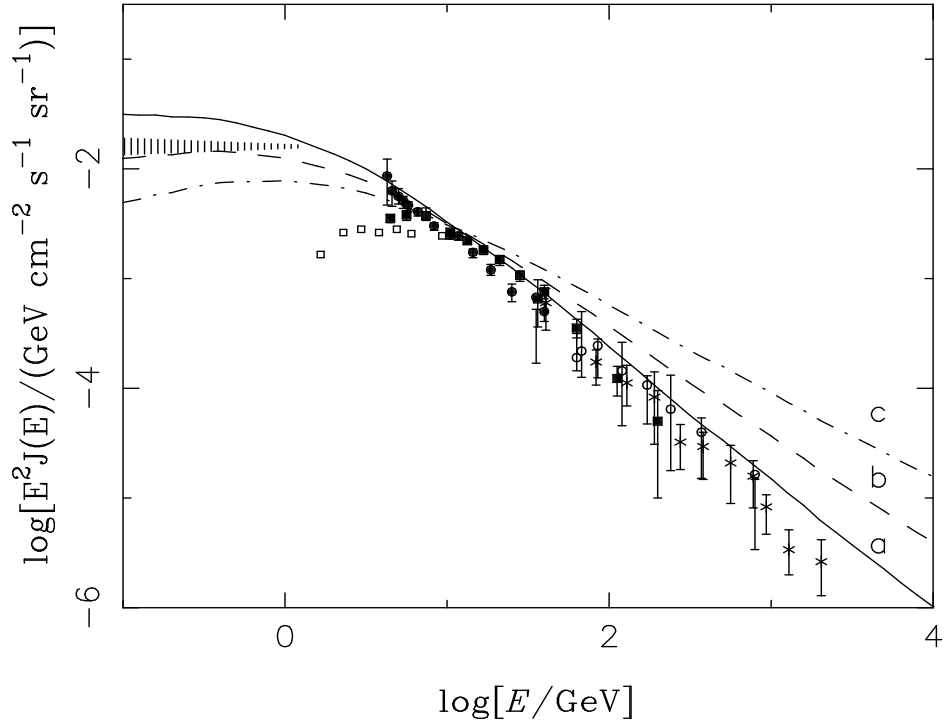}
\caption{Average interstellar electron spectrum normalised to observations
at 10 GeV \protect\cite{Tang84A,Golden94A,Nishimura95A} for an injection 
spectrum (a) E$^{-2.4}$, (b) E$^{-2.2}$ and (c) E$^{-2}$. 
Data points: taken from results summarised by Golden {\it et al} 
\protect\cite{Golden94A} and Nishimura {\it et al} 
\protect\cite{Nishimura95A}; Hatched band: electron spectrum derived
from non-thermal radio studies by Webber {\it et al}
\protect\cite{Webber80A}. }
\label{figESPECLoc}
\end{figure}

The interstellar electron spectrum is obtained from the probability matrix
and the source spectrum using Equation \ref{equNumMeth3}.
In Figure \ref{figESPECLoc} we show a comparison of direct measurements of the
electron intensity spectrum in the solar vicinity, spectra derived from 
non-thermal radio measurements and our predicted interstellar 
electron spectra with injection spectral indices $\gamma = 2$, 2.2 and 2.4
normalised to the observed spectrum at 10 GeV 
\cite{Tang84A,Golden94A,Nishimura95A} because for energies 
$\sim 10$ GeV and above 
solar modulation effects are believed to give only minor deviations 
from the interstellar spectrum, and the error bars on the measurements 
around 10 GeV are relatively small when compared with data at higher energies.
We can see that a flat injection index such as $\gamma = 2$ produces a 
local electron spectrum that neither agrees with the low energy spectrum
derived from non-thermal radio measurements, nor with the high energy 
data above 20 GeV. 
A flat injection spectrum such as $\gamma = 2$ is therefore, at least for 
the local region, ruled out, and we do not consider $\gamma = 2$ any 
further except for a special case which is discussed in 
Section \ref{subsectionExplan}.
Our predictions for $\gamma = 2.2$ and 2.4 are reasonably consistent 
with the data from 100 MeV up to roughly 50-60 GeV, with $\gamma = 2.2$ 
being slightly 
preferred by the spectrum derived from non-thermal radio observations.
At higher energies both spectra gradually over-predict the data such that
at 2 TeV a $\gamma = 2.2$ spectrum is a factor of 3 higher and a 
$\gamma = 2.4$ spectrum is a factor $\sim 1.5$ higher.
However we have assumed a continuous source 
distribution at all energies, and this is unlikely to be valid locally
as discussed below.

Electrons are assumed to be accelerated at strong supernova shocks, and the
nearest source region may be $\sim 100$ pc or more away.
At energies where the energy-loss time scale is less than or comparable
to the diffusion time to the nearest source high energy electrons will
lose most of their energy before reaching the observer.
The local spectrum will then be significantly steeper than the average 
interstellar spectrum \cite{Protheroe80A}.
Detailed calculations by Aharonian {\em et al} \cite{Aharonian95A} and 
Atoyan {\em et al} \cite{Atoyan95A} have shown that such a steepening 
occurs when the inhomogeneity of the source distribution is taken into
account.
However for X-rays and gamma-rays produced by electrons interacting with the 
galactic matter distribution, radiation and magnetic fields, the spectrum
of radiation observable at Earth depends on many contributions by the 
various X-ray and gamma-ray production processes along the line-of-sight.
This samples the spectrum of X-rays and gamma-rays, and hence the spectrum 
of electrons producing them, over large scales, and effectively
smooths out local inhomogeneities in the electron spectrum that could 
arise due to an uneven source distribution.
Therefore, at least in a first analysis, a continuous distribution of 
sources is reasonable when discussing gamma-ray observations. 

\begin{figure}[htb]
\vspace{15.0cm}
\includegraphics{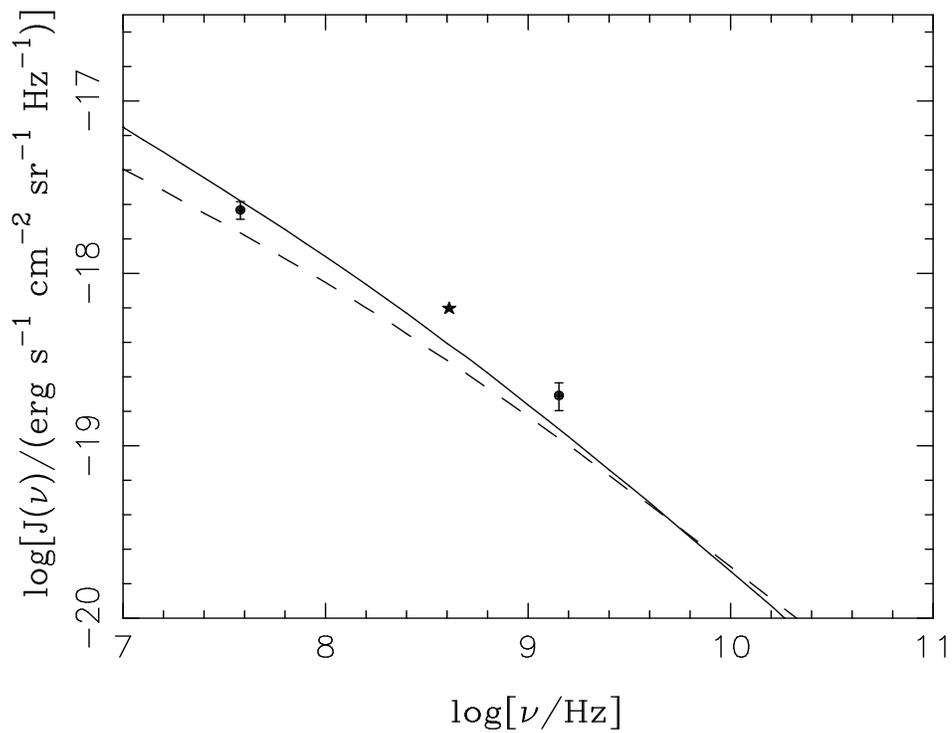}
\caption{Non-thermal radio emission in the direction of the galactic pole.
Solid curve shows the predicted emission for an $E^{-2.4}$ injection 
spectrum; dashed curve shows the predicted emission for an $E^{-2.2}$ 
spectrum. 
Data point at 408 MHz is from Broadbent {\it et al} 
\protect\cite{Broadbent90A} for a brightness temperature of 12.3 K. Results
at 38 MHz \protect\cite{Lawson87A} and 1420 MHz \protect\cite{Reich88A}
have been derived for synchrotron emission spectral index ranges 
$\alpha = 0.5 - 0.6$ and $\alpha = 0.85 - 1.1$ respectively.}
\label{figPolarRadioSpec}
\end{figure}

As a consistency check on our electron spectra, we calculate the non-thermal
emission in the direction of the galactic pole. We compare our predictions
with the results of Broadbent {\it et al} \cite{Broadbent90A}, who derive
a galactic pole brightness temperature of 12.3 K from modeling the 
408 MHz sky maps of Haslam {\it et al} \cite{Haslam82A}, and with 
the results of Lawson {\it et al} \cite{Lawson87A} at 38 MHz and 
Reich and Reich \cite{Reich88A} at 1420 MHz.
We show our predicted non-thermal spectrum at the pole in Figure 
\ref{figPolarRadioSpec} together with the experimental results.
Our results are within a factor of 1.5 of the data, indicating our 
electron spectra and magnetic field distribution are not unreasonable.
We use our calculated electron spectra in the next Section to obtain 
X-ray and gamma-ray emissivities for synchrotron, bremsstrahlung and IC 
radiation, and use these to compute diffuse galactic radiation spectra 
and compare with experimental results.

\section{Diffuse Galactic Gamma-Ray Spectra}
\label{sectionGammaSpec}

\subsection{Diffuse Gamma-Ray Emissivities}
\label{subsectionEm}
Our first step is to calculate the emissivity spectra of gamma radiation 
due to electron interactions with the interstellar medium, radiation and 
magnetic field. 
We calculate the IC, bremsstrahlung and 
synchrotron contributions to the diffuse gamma radiation from low to very high
energies. 
Emissivity spectra for the three processes are calculated using

\begin{equation}
q_{_{\rm Brem,IC,Sync}} \left( E_\gamma, \vec{r} \right) = \frac{4\pi}{c}
\int ^{E_{\rm max}} _{E_{\rm min}}
J \left( E, \vec{r} \right) \frac{d N_{\rm Brem,IC,Sync}}{dt dE_\gamma} dE
\hspace{5mm} ( {\rm GeV^{-1} \; cm^{-3} \; s^{-1}} )
\label{equDiffEmICBrem}
\end{equation}

\noindent
where $J ( E,\vec{r} )$ (GeV$^{-1}$ cm$^{-2}$ s$^{-1}$ sr$^{-1}$) 
is the interstellar electron spectrum, and 
$dN/dtdE_\gamma$ is the production spectrum for the appropriate process.
We calculate $dN/dtdE_\gamma$ for bremsstrahlung produced by 
electrons interacting with neutral and ionised hydrogen, and 
neutral helium. 
For IC gamma-rays $dN/dtdE_\gamma$ is evaluated using the Klein-Nishina
cross-section. 
Formulae for calculating $dN/dtdE_\gamma$ and the 
appropriate integration limits are given by 
Blumenthal and Gould \cite{Blumenthal70A} for bremsstrahlung and IC 
radiation, and by Pacholczyk \cite{Pacholczyk70A} for synchrotron radiation.

\begin{figure}[htb]
\vspace{15.0cm}
\includegraphics{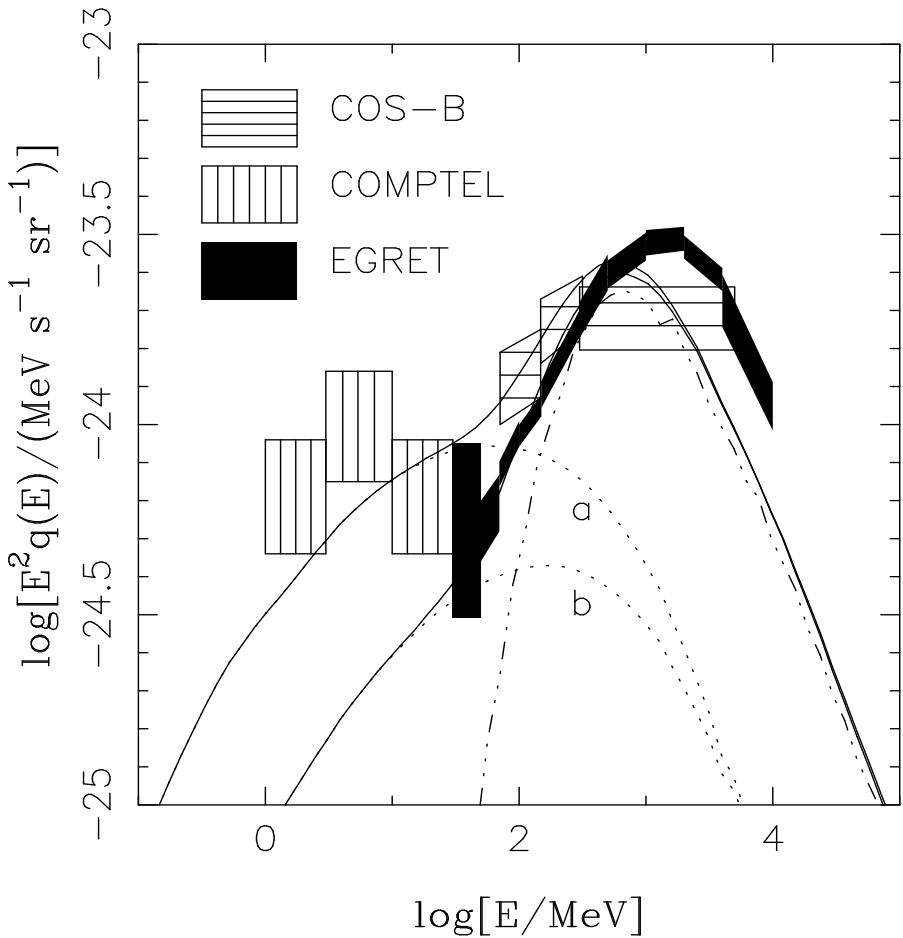}
\caption{Local emissivity spectrum of gamma-rays per interstellar H atom due
to matter interactions. 
Dotted curves show bremsstrahlung emissivity spectra
calculated for (a) $E^{-2.4}$ and (b) $E^{-2.2}$ injection spectra; chain
curve shows the expected $\pi^0$-decay emissivity as calculated by Stecker
\protect\cite{Stecker88A} and parameterised by Bertsch {\it et al} 
\protect\cite{Bertsch93A}. 
Observational data are taken from Strong and Mattox
\protect\cite{Strong96A}.}
\label{figBREMSPECLoc}
\end{figure}

Figure \ref{figBREMSPECLoc} shows the local emissivity per interstellar
H atom due to interactions with matter. 
The dotted curves show our results for bremsstrahlung for the two injection
spectra, and the chain curve shows the expected contribution from $\pi^0$-decay
according to Stecker \cite{Stecker88A} as parameterised by Bertsch {\it et al}
\cite{Bertsch93A}.
We also show the local emissivity per interstellar H atom estimated from
EGRET data (Strong and Mattox \cite{Strong96A}), COMPTEL data (Strong 
{\it et al} \cite{Strong96B,Strong94A}) and COS-B data (Strong {\it et al}
\cite{Strong88A}). 
A reasonable fit to the low energy data is obtained for an injection index
$\gamma = 2.4$, with the gamma-ray emissivities for $\gamma = 2.2$ 
being a factor of $\sim 2$ too low when compared with the data.
There is, however, some uncertainty in the COMPTEL spectrum amounting to 
a possible factor of two over-estimation \cite{Strong97A}.
With this taken into account a $\gamma = 2.2$ spectrum is allowed by the 
data.
As noted in several other papers \cite{Strong96B,Hunter97A,Strong96A}, above
500 MeV up to 10 GeV, the predicted emissivity due mainly to $\pi^0$-decay 
(not calculated in this paper), while consistent with the COS-B data, is 
significantly below the EGRET data.

\begin{figure}[htb]
\vspace{12.0cm}
\includegraphics{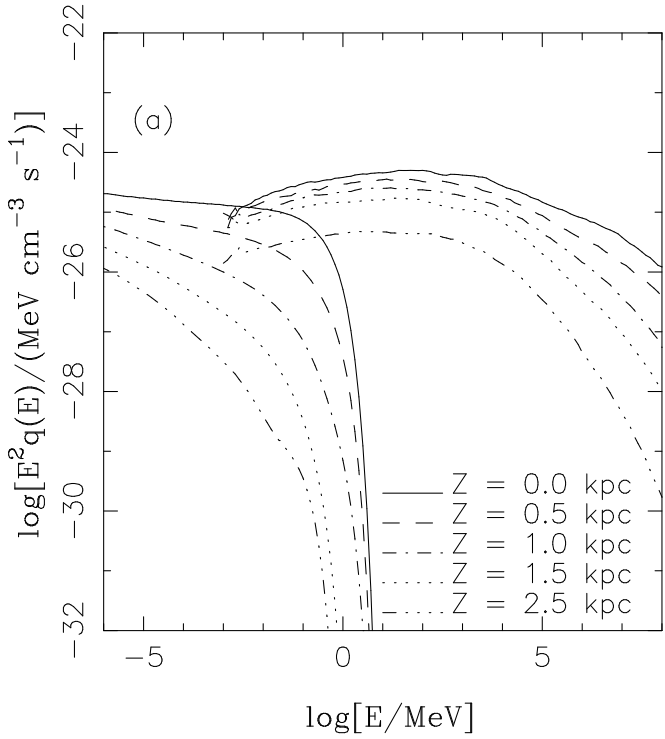}
\includegraphics{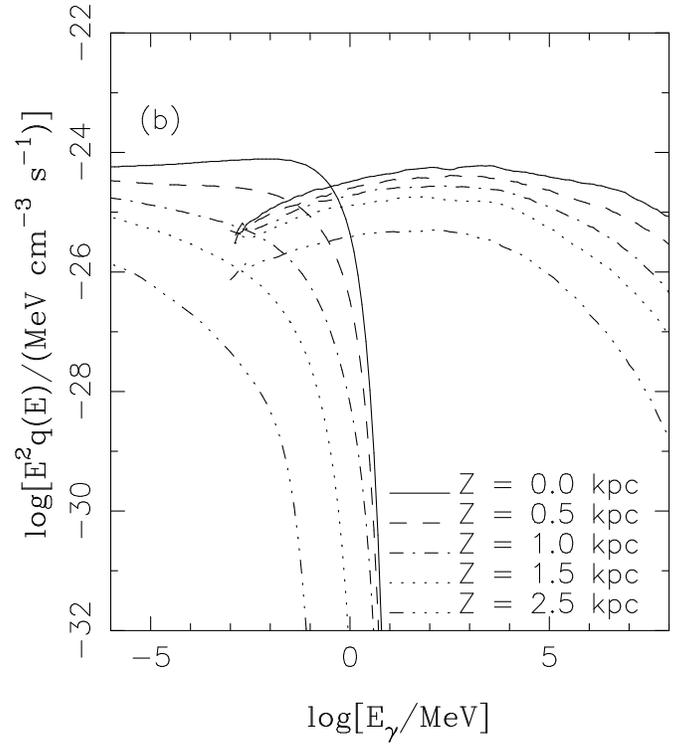}
\caption{Variation of the synchrotron and IC emissivity with 
height above the Galactic plane for the radial bin centred on $R = 9$ kpc 
and an injection spectrum of (a) $E^{-2.4}$ and (b) $E^{-2.2}$. 
Emission
spectra have been calculated using injection spectra with a maximum 
injection energy cut-off at 1 PeV.}
\label{figICSYNCSPECLoc}
\end{figure}

Figure \ref{figICSYNCSPECLoc} shows the variation with height above the 
galactic plane of the IC and synchrotron emissivities for 
the local region calculated for a high energy cut-off in the injection 
spectrum at 1 PeV; a cut-off in the injection spectrum of electrons at 
100 TeV to 1 PeV would be expected if electrons are accelerated by 
SNR \cite{Lagage83A,Hillas84A,Volk87A}.
Self-consistency for
both the IC and synchrotron contributions is ensured by using the
same galactic radiation field and magnetic field models for the emissivity 
calculation as used in the electron propagation. 
As can be seen in the Figure, synchrotron radiation can contribute 
significantly to the emission in hard X-rays, being at least comparable to 
the IC emission for energies $< 100$ keV in the galactic plane; at 
greater heights above the galactic plane the synchrotron spectrum decreases
more rapidly than the IC spectrum because the distribution of very high
energy electrons producing the synchrotron emission is significantly diminished
away from the plane, whereas the distribution of relatively low energy
electrons producing the IC spectrum extends to a considerable height above 
the plane. 
Variation of the injection spectrum index from $\gamma = 2.4$ to 2.2 
increases the synchrotron emission, and additionally tends to flatten the 
IC spectrum at high energies.
Furthermore, the cut-off in the synchrotron spectrum is  
directly linked to the cut-off in the injection spectrum of electrons:
a lower cut-off in the electron injection spectrum gives a 
correspondingly lower cut-off in the synchrotron spectrum.
This raises the interesting possibility of detecting signatures of maximum
acceleration energies of electrons using the galactic diffuse hard X-ray
background.

\subsection{Diffuse Gamma-Ray Intensities}
\label{subsectionInten}
\begin{figure}[htb]
\vspace{12.0cm}
\includegraphics{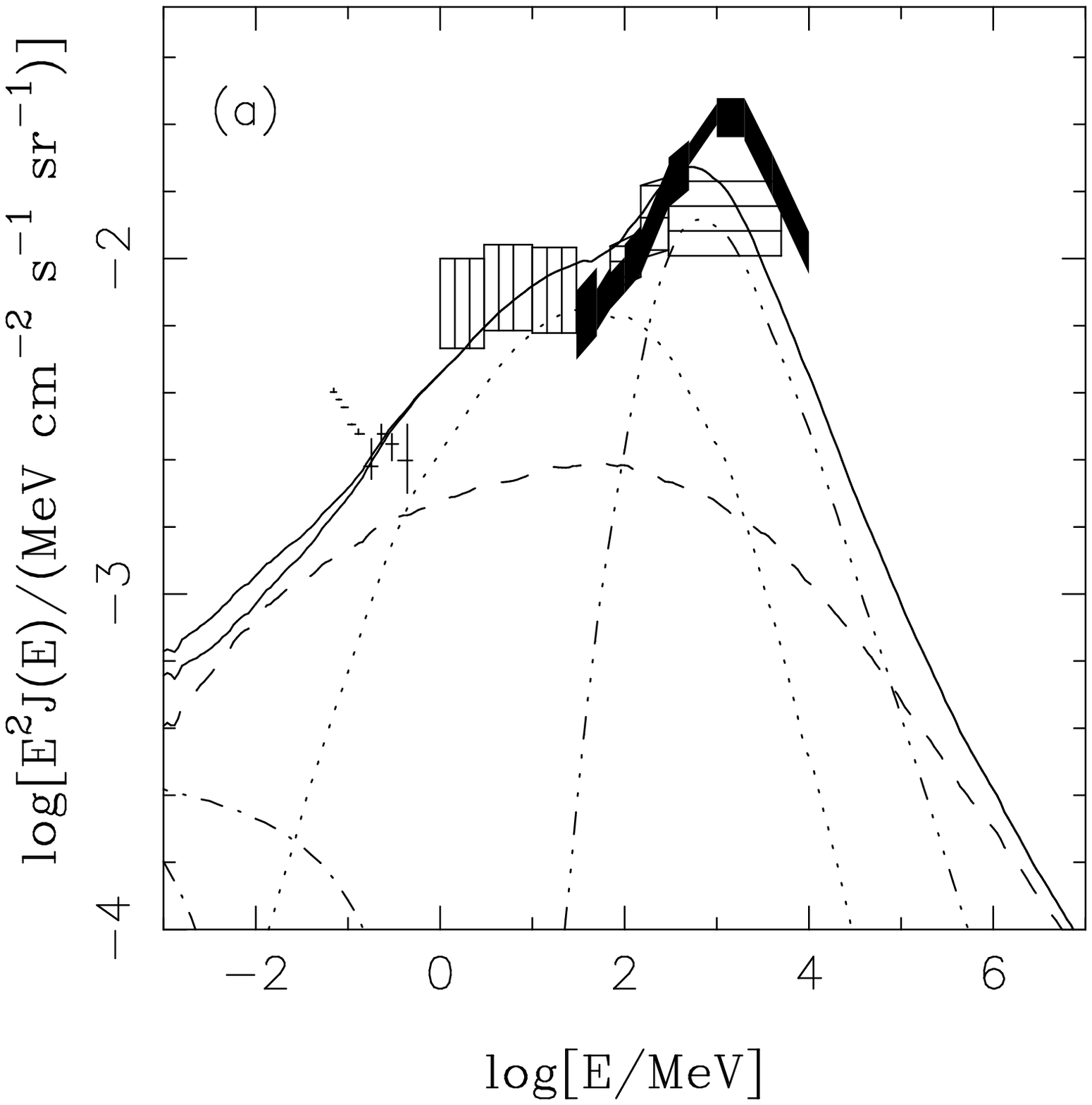}
\includegraphics{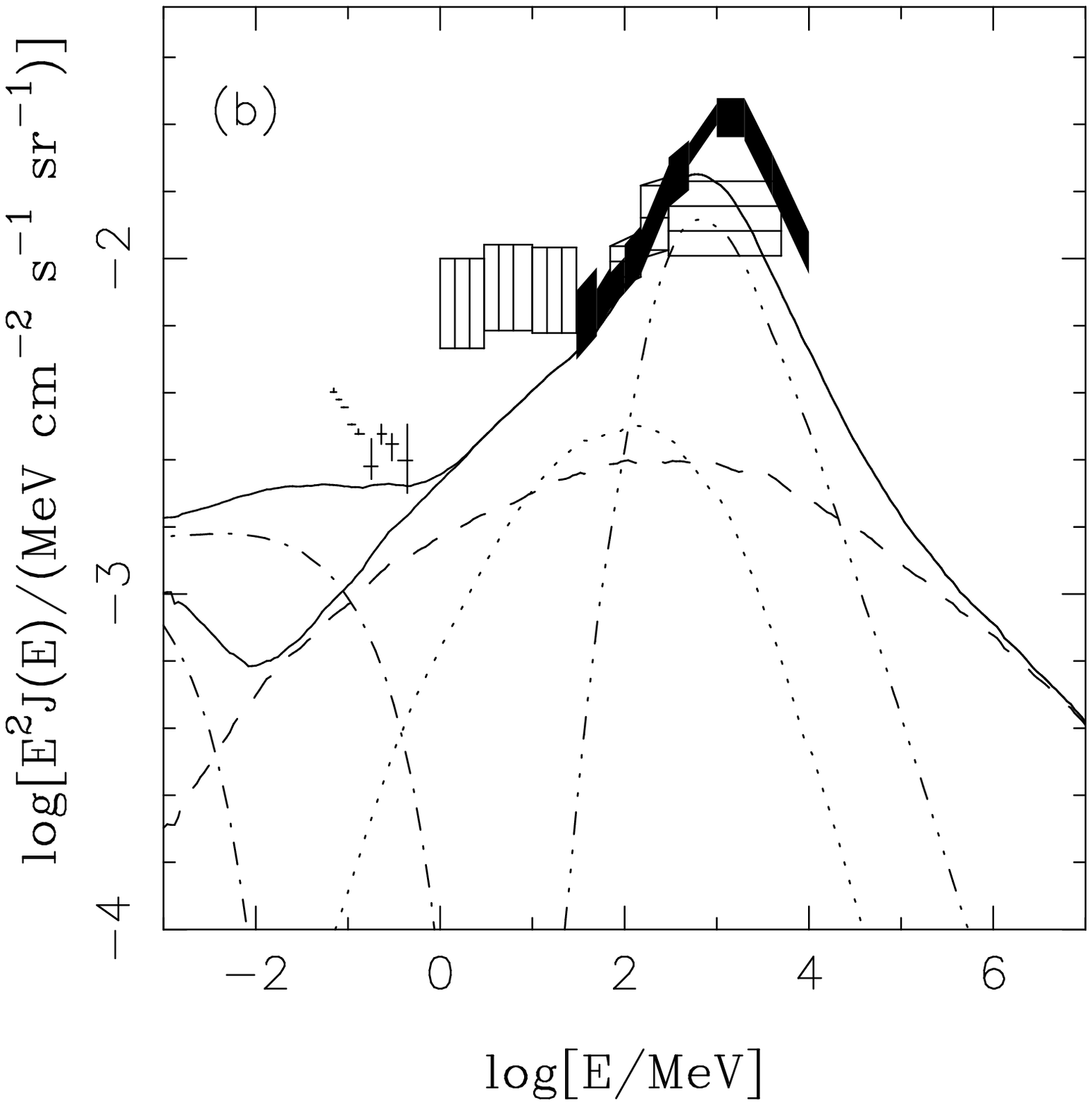}
\caption{Average gamma-ray spectra for the inner Galaxy ($-60^\circ < l <
60^\circ$ and $|b| < 20^\circ$) for an injection spectrum of (a) $E^{-2.4}$
and (b) $E^{-2.2}$. The individual contributions to the 
diffuse gamma-ray spectrum are indicated: IC - dashed curve;
bremsstrahlung - dotted curve; synchrotron - chain curve; $\pi^0$-decay -
double chain curve. The solid line is the sum of all contributions. 
Data 
are from various satellite telescopes; blocked data: EGRET 
\protect\cite{Strong96A}, horizontally hatched boxes: COS-B 
\protect\cite{Strong88A}, vertically hatched boxes: COMPTEL 
\protect\cite{Strong96B}, and data points: OSSE \protect\cite{Strong96B}
(original data from \protect\cite{Purcell95A}). }
\label{figINNERSpec}
\end{figure}

We use the emissivities calculated in the previous section to predict 
the intensity spectra for various gamma-ray production mechanisms, and compare
with observation. 
We calculate the average spectrum for the inner 
Galaxy, and show this in Figure \ref{figINNERSpec} for the two injection
spectra. 
Also shown are data from COS-B \cite{Strong88A}, COMPTEL 
\cite{Strong96B}, EGRET \cite{Strong96A} and OSSE \cite{Strong96B,Purcell95A}.
For the bremsstrahlung and $\pi^0$-decay spectra we have used the HI and CO
gas survey data from \cite{Strong88A} to obtain the average intensity. For
the synchrotron intensities we have used electron spectra with cut-offs of the
maximum injection energy at 100 TeV and 1 PeV respectively. 
The EGRET excess above 500 MeV apparent in the emissivity spectrum is also
present in the intensity spectrum, and our calculations indicate 
no simple modification in the electron spectrum can account for this 
feature. 
Also it should be
borne in mind that the uncertainty in the COMPTEL emissivity spectrum
mentioned in the previous section applies also to the COMPTEL intensity 
spectrum, i.e. it may be a factor $\sim 2$ too high \cite{Strong97A}. 
It can
be seen that a $\gamma = 2.4$ electron spectrum produces an intensity 
spectrum which
is largely consistent with the satellite data above 1 MeV, however it is 
unable to 
provide an explanation for the much softer gamma-ray spectrum observed by OSSE.

The fit to the data above 1 MeV for $\gamma = 2.2$ is acceptable when
the COMPTEL uncertainty is taken into account.
In the hard X-ray range a flatter spectrum appears for a $\gamma = 2.2$ 
spectrum due to the synchrotron emission by very high energy electrons in 
the galactic magnetic field, and is dependent on the cut-off in the 
primary injection spectrum of electrons, as noted in the previous section.
For neither injection spectrum is the OSSE data well fitted, and it has been
suggested that a new population of low energy electrons emitting 
bremsstrahlung radiation is required to provide the observed turn-up in 
the hard X-ray spectrum \cite{Skibo96A,Purcell96A,Skibo95A}.
However, it was noted in Section \ref{subsectionEm} that flattening the 
electron injection index gave an enhanced synchrotron emission.
If, for example, the injection spectrum in the inner Galaxy was flatter than
that locally, a flatter interstellar electron spectrum would result, and this
in turn would give an enhanced synchrotron emission.
We return to this point in Section \ref{subsectionExplan} 
when contrasting the hard X-ray and IC predictions.

At higher energies a dominant contribution by the 
IC component is of interest because the IC spectrum is 
directly related to the interstellar electron spectrum, which is itself 
directly related to the electron injection spectrum. 
At sufficiently high energies where the
IC spectrum could be unambiguously detected, information
about the electron injection spectrum can be obtained.
The Figure indicates the IC emission in the inner Galaxy is dominant 
above 30 GeV for $\gamma = 2.2$ and above 100 GeV for $\gamma = 2.4$.

\begin{figure}[htb]
\vspace{15.0cm}
\includegraphics{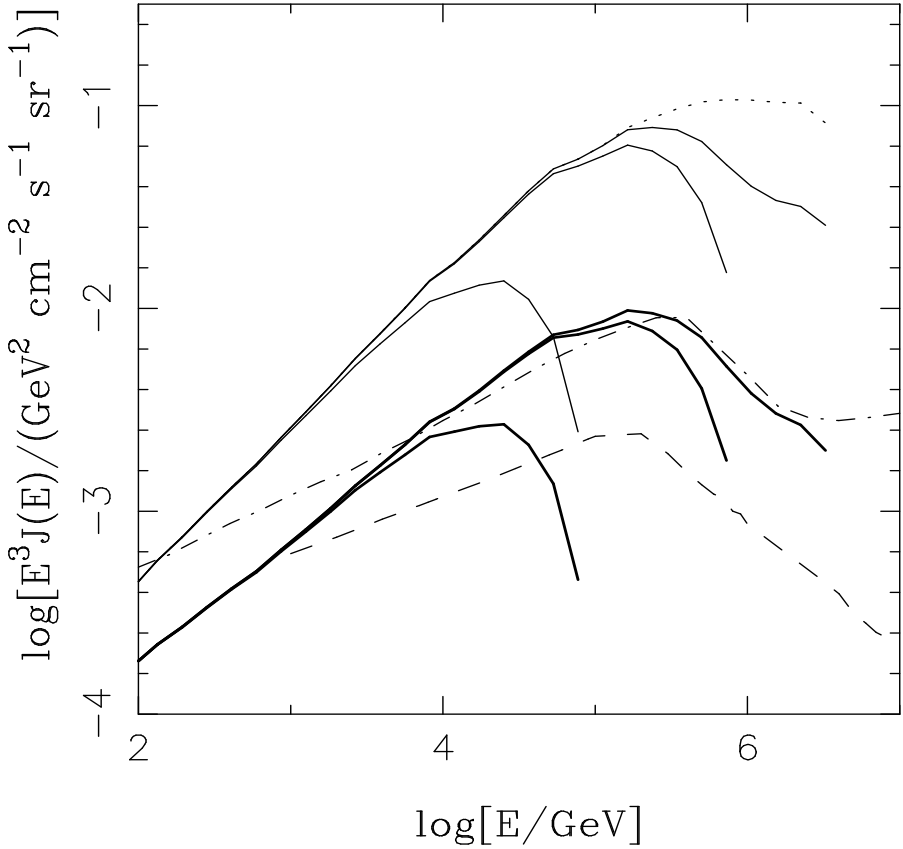}
\caption{Diffuse gamma-ray spectra in the direction $l = 0^\circ$, 
$b = 0^\circ$. 
Thick solid curves show the IC spectrum for 
an $E^{-2.4}$ injection spectrum of electrons; thin solid curves show the
IC spectrum for $E^{-2.2}$. 
For each injection spectrum, the lowest branch
is for a cut-off at 100 TeV, the next higher branch a cut-off at 1 PeV, and
the next higher no cut-off in the injection spectrum; each of these curves
includes attenuation on the CMBR. 
Dotted curve shows the IC spectrum for
an $E^{-2.2}$ spectrum with no cut-off and no attenuation on the CMBR.
Dot-dashed curve shows the predicted spectrum for $\pi^0$-decay (including 
attenuation on the CMBR) calculated by 
Ingelman and Thunman
\protect\cite{Ingelman96A}; Dashed curve shows the 
predicted $\pi^0$-decay spectrum (including attenuation on the CMBR) 
calculated by Berezinsky {\it et al} \protect\cite{Berezinsky93A}.}
\label{figGCSpec}
\end{figure}

We have computed the spectrum in the direction $l = 0^\circ$,
$b = 0^\circ$, and compared our predictions with recent calculations 
of the spectrum due to $\pi^0$-decay in the same direction 
\cite{Berezinsky93A,Ingelman96A}. 
We show the comparison in Figure \ref{figGCSpec} where we have calculated
IC spectra for $\gamma = 2.4$ and $2.2$, and for high 
energy cut-offs of the injection spectrum at 100 TeV and 1 PeV, and with  
no cut-off.
At energies above $\sim 100$ TeV the path-length of gamma-rays against 
photon-photon pair production on the CMBR is of the order size of the 
Galaxy. 
We therefore include attenuation on the CMBR when calculating our spectra; the
difference between the attenuated and unattenuated spectra is shown
in the top two branches of the $\gamma = 2.2$ curve in the Figure.
For energies in the range 1 MeV to 1 TeV the contribution by the 
IC process is predominantly due to scattering of optical and 
far infra-red photons. 
The optical component of the ISRF is the most uncertain due to the 
approximations used in the absorption calculation, and with estimates of 
the total 
luminosity varying by a factor of two \cite{Chi91A,Youssefi91A}.
Furthermore, we do not attempt a detailed modelling of the ISRF in the 
wavelength range 8 $\mu$m to 50 $\mu$m, however this region of the ISRF
is not as intense as the spectrum at shorter and longer wavelengths, and
therefore contributes a comparatively small amount to the total energy 
density of the ISRF. 
Given the uncertainties in the optical component, and our approximation 
of the middle infra-red spectrum, our predictions for the 1 MeV to 1 TeV 
energy range will be accurate only within a factor of two.
For higher energies, IC scattering of optical and far infra-red photons is 
in the Klein-Nishina regime, and so scattering of CMBR photons contributes
to the bulk of the spectrum.
Thus the uncertainty in our predictions due to uncertainties in the 
radiation field model diminishes at high energies.
In the Figure we see that a $\gamma = 2.4$ spectrum is at least 
comparable to the lowest of the $\pi^0$-decay predictions at 1 TeV (note
the differences between the two $\pi^0$-decay predictions are due, in part,
to the differences in the cosmic-ray spectrum and the pion multiplicities
used), and that a $\gamma = 2.2$ spectrum is about a factor of 3 higher 
than the $\gamma = 2.4$ prediction at 1 TeV, increasing slowly with increasing
energy. 
At the highest particle energies in our calculations the IC
spectrum is at least of a similar magnitude to the largest $\pi^0$-decay
predictions.
Hence the gamma-ray intensity above 1 TeV can have a significant contribution
due to IC scattering.

\begin{figure}[htb]
\vspace{15.0cm}
\includegraphics{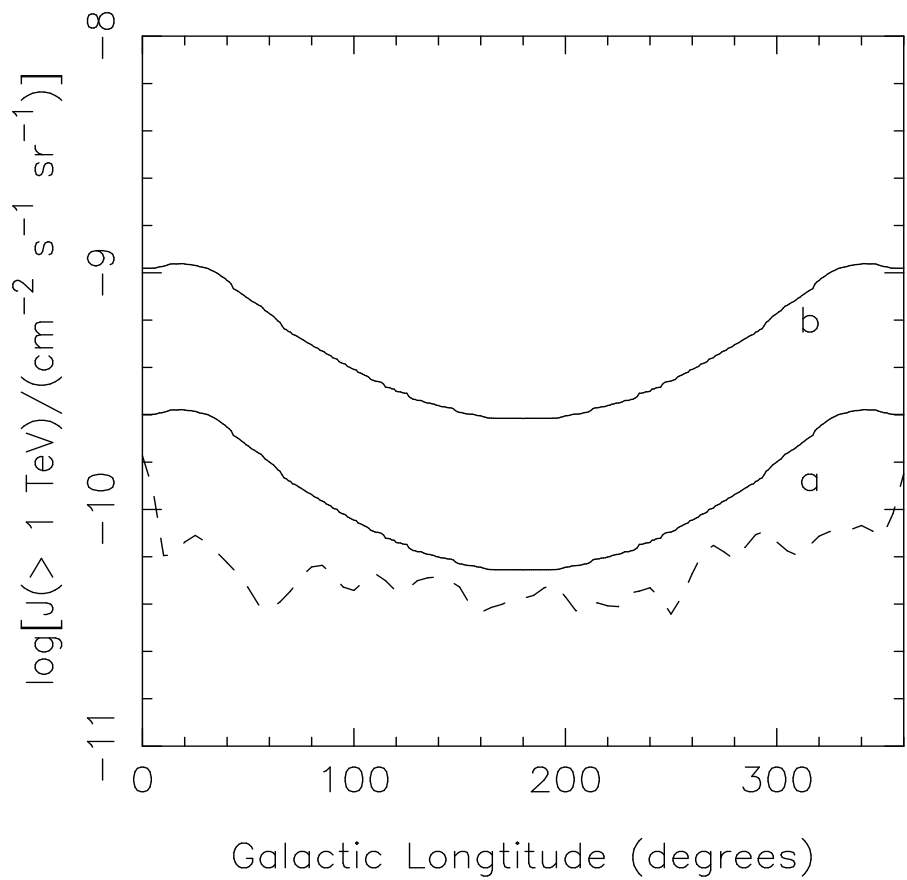}
\caption{Longitude profile of the IC intensity above 1 TeV
averaged over $|b| < 10^\circ$ for an injection spectrum of (a) $E^{-2.4}$
and (b) $E^{-2.2}$. 
Dashed curve shows the predictions of 
Berezinsky {\it et al} \protect\cite{Berezinsky93A} for cosmic-ray 
interactions with matter in the Galaxy.}
\label{figPROFILE1TeV}
\end{figure}

We calculate a longitude profile for the integral gamma-ray spectrum
above 1 TeV and show this in Figure \ref{figPROFILE1TeV} together with the
expected profile of $\pi^0$-decay gamma-rays as predicted by Berezinsky 
{\it et al} \cite{Berezinsky93A}.
It can be seen the TeV gamma-ray intensity is sensitive to the 
electron injection spectrum.
Additionally, introducing a high energy cut-off in the injection spectrum
results in a well defined high energy cut-off in the IC
spectrum as in Figure \ref{figGCSpec}.
Hence measurements of the diffuse gamma-ray spectrum beyond TeV energies
may provide a direct means of estimating the electron source spectrum.

Upper limits on the diffuse gamma-ray spectrum for energies $\sim 50 - 500$ 
TeV have only been published for experiments located in the Northern 
Hemisphere.
Observations of the galactic centre region are precluded by the location
on Earth of these experiments.
At $\sim 400$ GeV energies the Whipple group \cite{Reynolds93B} have
obtained upper limits on the ratio of the diffuse gamma-ray spectrum to the
all particle cosmic-ray spectrum.
At $\sim 50 - 500$ TeV energies the Utah-Michigan array \cite{Matthews91A}
and CASA-MIA \cite{Borione97A} have searched for diffuse 
emission in the
region of the sky corresponding to the galactic coordinates 
$50^\circ < l < 200^\circ$ and $|b| < 10^\circ$, and have obtained upper 
limits on the ratio of diffuse gamma-rays to total cosmic-ray intensity.
Published data for the HEGRA array \cite{Karle95A} covers a region of sky
that includes portions of the galactic plane as well as high galactic 
latitudes.

\begin{figure}[htb]
\vspace{17.0cm}
\includegraphics{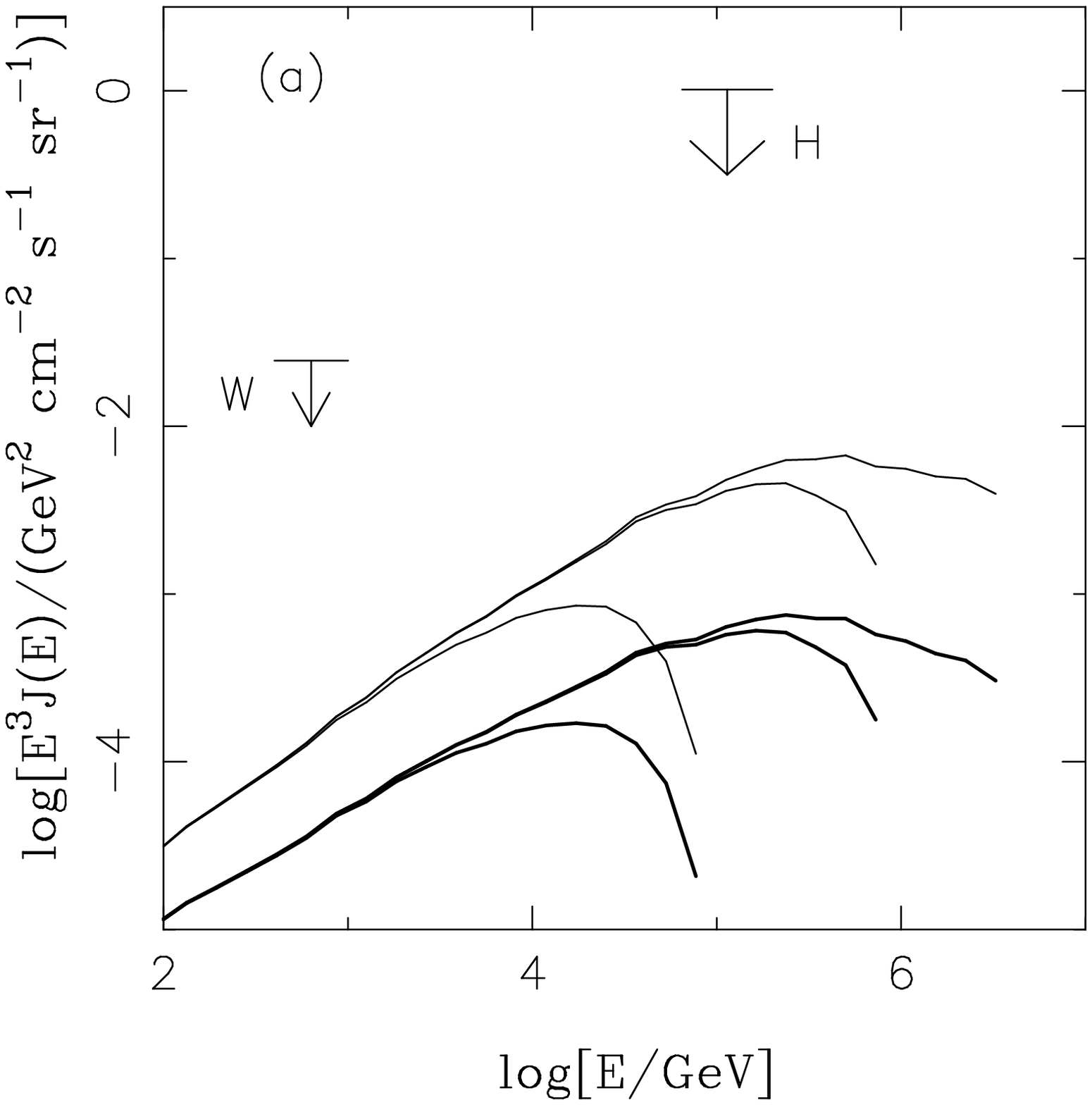}
\includegraphics{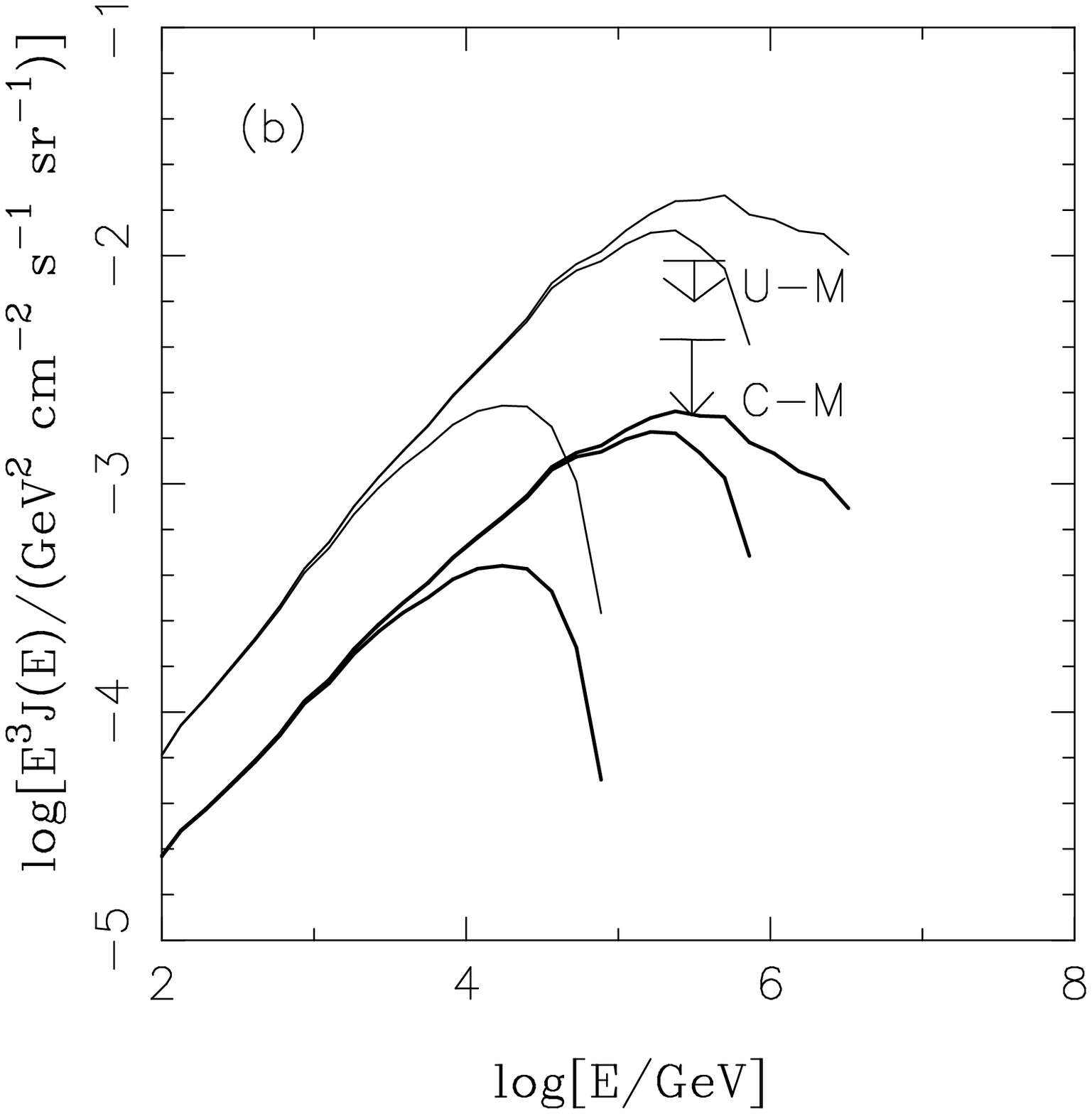}
\caption{Diffuse gamma-ray spectra due to IC interactions 
predicted for the injection spectra $E^{-2.4}$ (thick solid curve) and 
$E^{-2.2}$ (thin solid curve) averaged over the region 
of sky covered by (a) Whipple telescope (Reynolds {\it et al} 
\protect\cite{Reynolds93B}) and HEGRA (Karle {\it et al} 
\protect\cite{Karle95A}) 
and (b) the Utah-Michigan array (Matthews {\it et al} 
\protect\cite{Matthews91A}) and CASA-MIA (Borione {\it et al} 
\protect\cite{Borione97A}).
The lower branch of each curve corresponds to a cut-off in the injection 
spectrum at 100 TeV, the next highest 1 PeV, and the highest branch no
cut-off; each of the curves includes attenuation on the CMBR.}
\label{figHEGRAandCASASpec}
\end{figure}

To make a proper comparison with the observational data, we have calculated 
the expected spectrum of diffuse IC 
radiation averaged over the region of the sky covered by the Utah-Michigan
array and CASA-MIA, and the region of sky covered by the HEGRA array (being
at a similar latitude, the Whipple Observatory covers a region of sky
similar to that of the HEGRA array).
We show our predictions in Figure \ref{figHEGRAandCASASpec} 
for the the two injection indices, together with cut-offs in the 
injection spectrum at 100 TeV and 1 PeV respectively, along with the 
upper limits for the diffuse gamma-ray spectrum obtained by converting the
ratios $I_\gamma/I_{CR}$ from the above experiments to gamma-ray intensity
upper limits using the all particle cosmic-ray spectrum \cite{Stanev92A}.
The Whipple and HEGRA upper limits do not constrain either the injection
index, or the cut-off in the injection spectrum.
The CASA-MIA and Utah-Michigan upper limits, however, do at least allow
us to rule out several of the cases considered. For example, injection
energies beyond $\sim 500 - 700$ TeV are ruled out for a $\gamma = 2.2$ 
injection spectrum. 
For a $\gamma = 2.4$ spectrum no constraints are placed by the 
experimental results.

\subsection{Possible explanation for the steep hard X-ray spectrum}
\label{subsectionExplan}
We now contrast the very high energy IC 
predictions and the hard X-ray synchrotron results, and suggest a possible
explanation for the steep hard X-ray spectrum.
For an $E^{-2.4}$ injection spectrum, no constraints are placed by either the
hard X-ray nor the IC predictions, 
and below 1 MeV the spectrum is not of the required shape and gives a poor fit
to the data (see Figure \ref{figINNERSpec}).
For this case, a new component of the galactic background radiation is needed
at low energies to explain the turn up in the OSSE spectrum, and Skibo
{\em et al} \cite{Skibo96A} have argued that a new population of non-thermal
bremsstrahlung producing electrons with steep spectra ($\propto E^{-2.5}$)
are required to explain the observed spectrum.
Assuming the observed X-ray spectrum is truly diffuse, their approach 
suggests a power required to maintain the electrons
against energy losses in the interstellar medium about an order of magnitude
larger than that supplied by galactic supernovae \cite{Skibo96A,Skibo95A}.
On the other hand the hard X-ray spectrum may be due to the sum of 
a large number of steep power-law unresolved point sources in which case 
the problems with the energetics in the Skibo {\em et al} model are 
circumvented.
In any case an $E^{-2.4}$ spectrum is realistically only able to provide
a good fit to the data between 1 MeV and $\sim 500$ MeV, with no 
constraints provided by the high energy IC spectrum.

For an $E^{-2.2}$ spectrum, the high energy spectrum is constrained 
by the CASA-MIA data: no electrons with energies greater than $500 - 700$ TeV
can be accelerated.
The hard X-ray predictions give no indication of a maximum injection 
energy cut-off, and at first glance the fit to the OSSE data seems not
good enough to suggest a synchrotron origin for the hard X-ray spectrum.
However in Section \ref{subsectionEm} we noted that as the injection 
spectrum flattened from $\gamma = 2.4$ to 2.2 the magnitude of the synchrotron 
emission became correspondingly greater.
If the injection spectrum was flattened further to, for example, 
$\gamma = 2$ then the OSSE data may be able to be fitted.
As our calculations in Section \ref{subsectionCalc} indicated, such an
injection spectrum gives an interstellar electron spectrum inconsistent 
with the local one, and this would seem to be not allowed.
This assumes a single injection index adequately describes electron
acceleration over the entire Galaxy.
However, toward the inner Galaxy, in regions of greater star formation, 
there could be a large population of unresolved young SNR with relatively
flat injection spectra, and these could give the required spectrum  
toward the inner Galaxy; it must be borne in mind that the CASA-MIA results
are for the galactic disk $50^\circ < l < 200^\circ$ which  
excludes most of the inner region of the Galaxy.
Towards the outer Galaxy the star formation rate is not as great, and the 
SNR located in this region are sparser and there are probably fewer in the
early stages off 
their evolution, and hence most are less efficient accelerators of 
electrons \cite{Drury94A}.
This could give a steeper injection spectrum towards the outer Galaxy than
the inner Galaxy.

\begin{figure}[htb]
\vspace{12.0cm}
\includegraphics{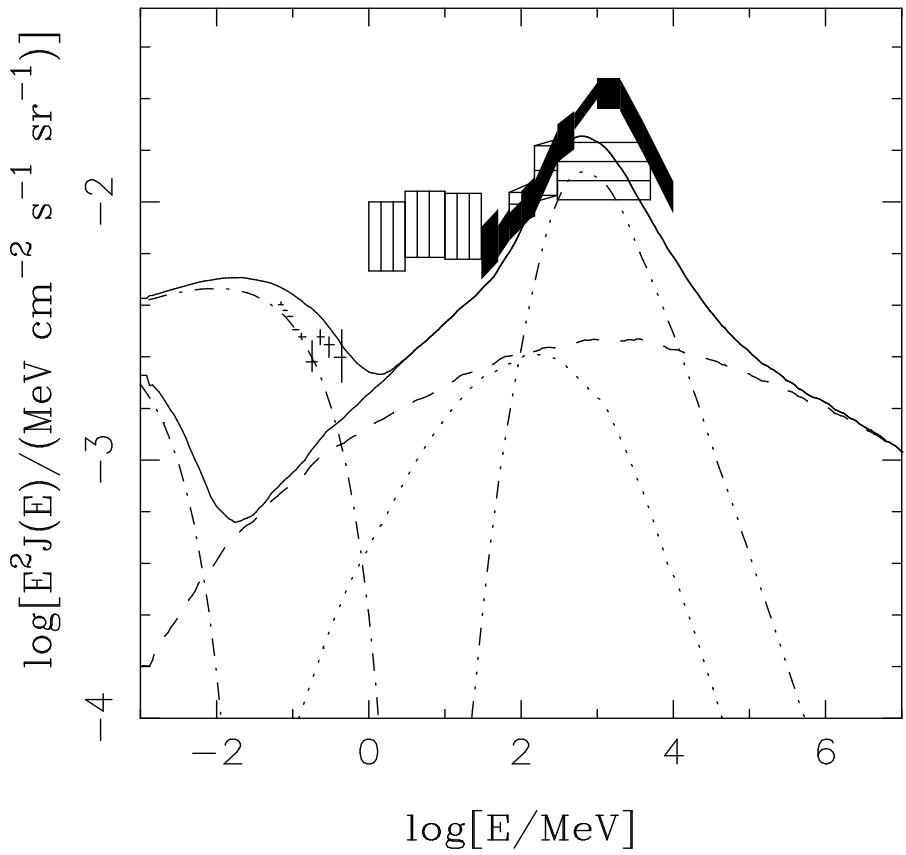}
\caption{Average gamma-ray spectra calculated using the modified source 
model described in the text for the inner Galaxy; curves have the
same meaning as in Figure \protect\ref{figINNERSpec}.}
\label{figMODINJSpec}
\end{figure}

We have computed average gamma-ray spectra for both the inner Galaxy and 
the region of sky covered by CASA-MIA using a `modified' source model 
comprising of an $E^{-2}$ injection spectrum for $R < 6$ kpc, and an 
$E^{-2.2}$ injection spectrum for $R > 6$ kpc; we re-use the 
probability matrix for the appropriate radial bin together with the 
modified source model to obtain electron spectra for the Galaxy, as
described in Section \ref{subsectionCalc}.
In the absence of truly three-dimensional diffusion in our calculations we
obtain the normalisation for the electron spectra in the inner Galaxy 
by normalising to the local data around 10 GeV, and then scaling 
according to the distribution of SNR from Section \ref{subsectionCalc}
; this, at best, approximate method at least allows us to demonstrate how a
fit to the gamma-ray spectra might be obtained.
We show our prediction for the inner region of the Galaxy using 
this combination of injection spectra in Figure \ref{figMODINJSpec}.
The fit to the diffuse hard X-ray to gamma-ray spectrum is greatly improved 
(bearing in mind the possible COMPTEL uncertainty) and could be made to 
give an even better fit with minor changes to the model input. 
For the outer region of the Galaxy 
(not plotted) we obtain a high energy spectrum which is almost identical
to the $\gamma = 2.2$ spectrum in Figure \ref{figHEGRAandCASASpec}b.
It is interesting to note that it has been suggested by Hunter {\it et al}
\cite{Hunter97A} that a possible 
explanation for the EGRET excess above 500 MeV is that the spectrum of 
protons is flatter in the inner Galaxy than the outer Galaxy, 
reflecting a fairly flat source spectrum $\propto E^{-( 2 - 2.3)}$ toward
the inner Galaxy.
If protons and electrons are accelerated with the same power-law then this
possible explanation for the anomalous EGRET results would suggest a 
flatter electron 
source spectrum in the inner Galaxy, which corresponds well with our 
suggestion for the origin of the steep OSSE spectrum.

\section{Discussion and Summary}
\label{sectionDiscuss}

We discuss the limitations of our predictions and summarise.
The main limitation in our calculation is the one-dimensional approximation
used for the electron propagation: it effectively ignores particle 
density gradients due to radial diffusion, and does not allow us to 
adequately take into account the inhomogeneous distribution of sources in 
the Galaxy.
Uncertainties exist in our diffuse gamma-ray predictions above
1 MeV and below 1 TeV due to the radiation field model used, however 
these are relatively 
minor and do not significantly affect the results below 1 MeV and above 1 TeV
where electrons IC scatter predominantly CMBR photons.
The galactic magnetic field is not entirely well known, and the model 
we use, while adequate given the other approximations made in our 
calculations, is quite simple.
Features such as a general increase in the field strength 
toward the galactic centre \cite{Rand94A} are not included.
Taking such an increase into account would have interesting results for our 
hard X-ray and very high energy gamma-ray predictions, mainly because an 
increased field strength would give an increased synchrotron emissivity 
toward the inner Galaxy, and would also tend to steepen the high energy
electron spectrum with the increased synchrotron energy losses.
In future work we plan to address these points in greater detail.

The galactic background radiation has been observed at X-ray energies 
by satellites such as OSSE and Ginga \cite{Yamasaki96A}, and at MeV to GeV
energies by COMPTEL, COS-B and EGRET.
Beyond 400 GeV upper limits on the galactic background have been obtained
by optical $\check{\rm C}$erenkov telescopes and air shower arrays but, at
present, no connection exists between the energy ranges observed by 
satellite-borne and ground based detectors.
The proposed next generation of gamma-ray satellites, such as GLAST 
\cite{Bloom96A} and GAMMA-400 \cite{Fradkin95A}, will rectify this 
situation.
Both the GLAST and GAMMA-400 groups provide estimates of the sensitivity of 
their proposed instruments, and they should be able to improve by approximately
two orders of magnitude the Whipple upper limits in $\sim 1$ year of 
operation.
Improvements of this order would provide further constraints on the 
high energy diffuse gamma-ray spectrum, and hence the high energy interstellar
electron spectrum.

To summarise, we have calculated the spectrum of galactic background 
radiation from keV to TeV energies and above.
We find that an interstellar electron spectrum corresponding to 
an $E^{-2.2}$ source spectrum results in a high energy IC 
spectrum that dominates the diffuse gamma-ray spectrum from 30-50 GeV up to
1 PeV.
Furthermore, such a source spectrum results in a fairly flat diffuse hard X-ray
spectrum due to synchrotron radiation by high energy electrons in the
galactic magnetic field.
For a source spectrum $E^{-2.4}$ the high energy IC spectrum 
is comparable to recent estimates of the spectrum of gamma-rays from 
$\pi^0$-decay, and there appears no significant contribution to the hard
X-ray background for electrons accelerated with this spectrum.
Assuming that a single injection index adequately describes electron 
acceleration throughout the entire Galaxy, upper limits on the diffuse
gamma-ray flux by optical $\check{\rm C}$erenkov telescopes and air shower
arrays rule out acceleration of electrons with energies higher than 
$500 - 700$ TeV for an $E^{-2.2}$ source spectrum, but provide no
constraints for an $E^{-2.4}$ source spectrum.
For both injection spectra, no constraints are provided by the hard X-ray 
spectrum.
However, we have shown that it is possible to obtain a fairly good fit 
to the hard 
X-ray spectrum if cosmic-ray electrons are accelerated with a flatter
source spectrum in the inner Galaxy than the outer Galaxy.
This provides an alternative origin for the hard X-ray spectrum than 
has otherwise been proposed, and corresponds well with an explanation for 
the EGRET results above 500 MeV.
Future observations above 10 GeV with improved sensitivities can be used
to better measure the spectrum of diffuse galactic 
gamma-rays and, when combined with air shower observations, provide 
further constraints on the spectrum of cosmic-ray electrons at acceleration.

\section*{Acknowledgements}
We thank Jim Matthews for discussions concerning CASA-MIA, and 
Andrew Strong for clarifying certain aspects of COMPTEL measurements.
We thank Wlodek Bednarek for reading the manuscript.
This work was supported by the Australian Research Council.

\newpage

\end{document}